\documentclass{article}
\usepackage{spconf,amsmath,graphicx}

\usepackage{enumitem}
\setlist{nosep, leftmargin=14pt}
\usepackage{subfig}
\usepackage{mwe} 

\title{COVID-19 computer-aided diagnosis through AI-assisted CT imaging analysis: Deploying a medical AI system}
\name{Demetris Gerogiannis $^1$, Anastasios Arsenos $^2$, Dimitrios Kollias $^3$, Dimitris Nikitopoulos $^4$, Stefanos Kollias $^2$ $^4$}
\address
{$^1$ Dept. of Computer Science \& Engineering, University of Ioannina / AIandMe SMPC - Ioannina, Greece \\
$^2$ School of Electrical \& Computer Engineering, National Technical University of Athens, Greece \\
$^{3}$  School of Electronic Engineering \& Computer Science, Queen Mary University of London, UK \\
$^{4}$  GRNET, National Infrastructures for
Research \& Technology}

\begin{document}
%
\maketitle
\begin{abstract}
Computer-aided diagnosis (CAD) systems stand out as potent aids for physicians in identifying the novel Coronavirus Disease 2019 (COVID-19) through medical imaging modalities. In this paper, we showcase the integration and reliable and fast deployment of a state-of-the-art AI system designed to automatically analyze CT images, offering infection probability for the swift detection of COVID-19. The suggested system, comprising both classification and segmentation components, is anticipated to reduce physicians' detection time and enhance the overall efficiency of COVID-19 detection. We successfully surmounted various challenges, such as data discrepancy and anonymisation, testing the time-effectiveness of the model, and data security, enabling reliable and scalable deployment of the system on both cloud and edge environments. Additionally, our AI system  assigns a probability of infection to each 3D CT scan and enhances explainability through anchor set similarity, facilitating timely confirmation and segregation of infected patients by physicians.

\end{abstract}
\section{Introduction}
\label{sec:intro}

The increasing number of cases makes the manual examination of image modalities time-consuming. Introducing machine learning into the biomedical field can significantly support physicians in efficiently and effectively conducting computer-aided diagnoses of medical images. CAD systems play a crucial role in helping radiologists interpret medical images, and, therefore, computer-aided detection becomes a valuable tool for distinguishing radiographic images indicative of COVID-19 infection. The application of Artificial Neural Networks (ANNs), Machine Learning (ML) and Deep Learning (DL) in CAD systems has demonstrated remarkable success in the analysis of medical data \cite{kollias2020transparent, kollias2018deep}.

This paper introduces a Computer-aided diagnosis application built on a microservices approach and is based on a state-of-the-art deep learning model named RACNet \cite{kollias2023deep}. The key features of the proposed system are  effectiveness, data anonymisation and fairness, and enhanced explainability of the AI-assisted diagnosis. In particular, for training and evaluation of system a newly developed extensive database, COV19-CT-DB \cite{arsenos2022large, arsenos2023data}, was utilised comprising chest 3-D CT scans sourced from various hospitals. The database encompasses $7,756$ annotated 3-D CT scans, distinguishing between $1,661$ COVID-19 cases and $6,095$ non-COVID-19 cases. In addition, a DL architecture, RACNet, is designed to analyze 3-D CT scan inputs, effectively handle varying numbers of CT slices per scan, and deliver exceptional performance on COV19-CT-DB and other public datasets for COVID-19 diagnosis. RACNet, a CNN-RNN architecture, is enhanced with routing and feature alignment steps dynamically selecting specific RNN outputs for decision-making in COVID-19 diagnosis. Furthermore, latent variables from the trained RACNet are extracted, deriving a set of anchors that offer insights into the network's data-driven knowledge. 

As far as the application deployment concerned, a microservices \cite{soldani2018} architecture is introduced followed to build the end users application. Note that some critical functionalities related to the handling of sensitive information are executed on the edge (i.e., doctor's local environment) while the intense computational models are executed on a HPC cloud. The underlying proposed architecture facilitates the data flow and automates the communication process tackling also the security issues. In summary, the paper makes the following contributions:

\begin{enumerate}
    \item A robust system architecture for deploying scalable and secure AI applications on heterogeneous computational environments.
    \item The AI system proposed in this study has the potential to alleviate physician workload in outbreak regions by prioritizing disease cases and enhancing diagnostic accuracy.
    
\end{enumerate}
\section{Materials and Model}
\subsection{Dataset}

COV19-CT-DB, the dataset used for training the deep learning model of the proposed application, comprises annotated 3-D chest CT scans indicating the presence or absence of COVID-19. The data is anonymised, with multiple CT scan series often associated with each individual, typically captured at different time points. In total, the database includes $724,273$ slices for CT scans in the COVID-19 category and $1,775,727$ slices for the non-COVID-19 category. The dataset is partitioned into training, validation, and test sets. \
The diverse (data originating from different Hospitals) and representative (huge amounts of data) are essential for achieving data fairness in our application. This dataset  reflects the diversity of the real-world population. As a consequence, it enables the AI model to learn from a comprehensive set of examples, fostering equitable performance across different environments and minimizing the risk of discriminatory outcomes.

\subsection{Deep learning model}

\begin{figure} [h!]
\centering
\includegraphics[width=0.96\linewidth]{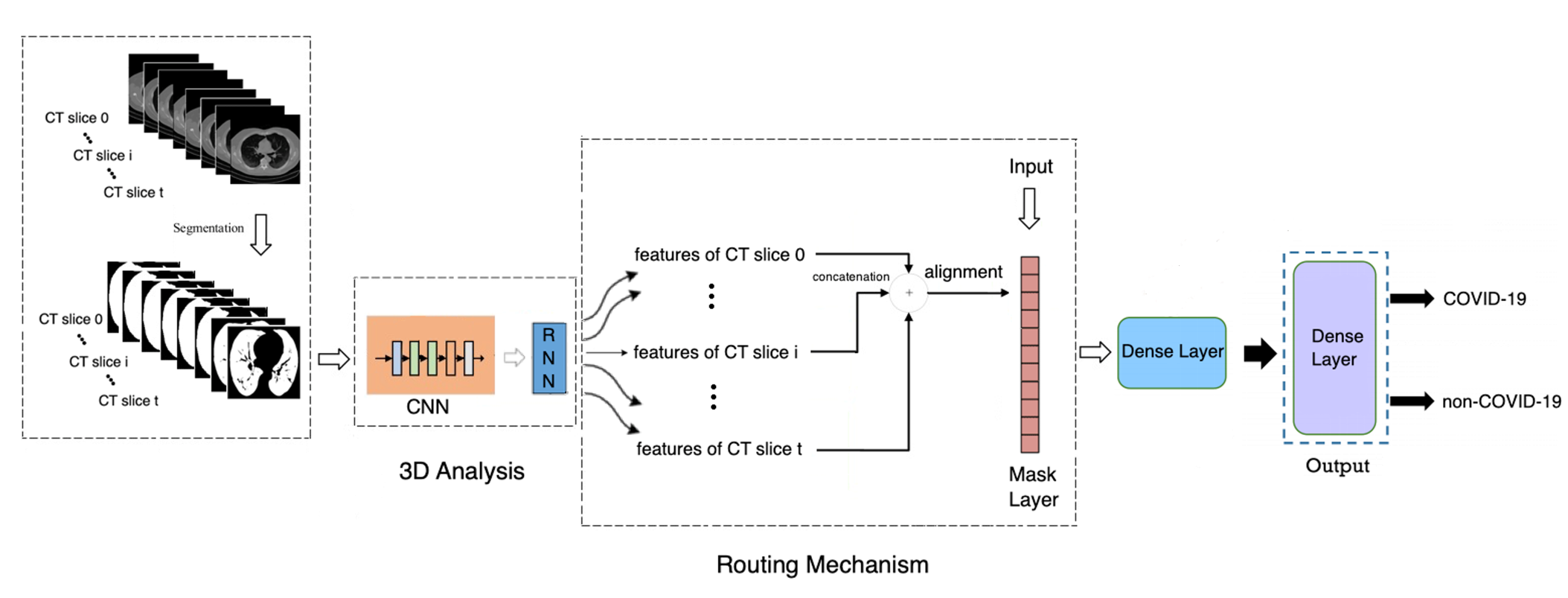}
\caption{RACNet architecture}
\label{fig:RACNet}
\end{figure}

In this paper, we use a CNN-RNN architecture, Routing-Align-ClusterNet (RACNet) \cite{kollias2023deep}. RACNet (Figure\ref{fig:RACNet}) consists of three components performing 3D Analysis, Routing and Classification respectively.

The input data undergoes processing in the 3D Analysis component of RACNet. During RACNet training, the routing mechanism dynamically selects RNN outputs/features. This component comprises a CNN network followed by an RNN. The CNN conducts localized analysis on a per 2-D slice basis, primarily extracting features from lung regions. It is noteworthy that the objective is to diagnose using the entire 3-D CT scan series, similar to the approach taken by medical experts during annotation. The RNN component facilitates this diagnosis by sequentially analyzing CNN features across the entire 3-D CT scan, progressing from slice $0$ to the last available slice $t - 1$. Classification of the selected RNN outputs is performed by a dense layer. This feeds the output layer, which provides a  positive, or negative, disease diagnosis decision.

In particular, the key advantages of RACNet compared to other approaches, such as 3-D CNNs, consist of the following: a) it harmonizes analysis of 3-D image volumes consisting of different number of slices and annotated per volume, b) it unifies decisions made over different datasets, thus enriching data-driven knowledge and improving its trusted use. RACNet was successfully used in COVID-19 diagnosis based on chest 3-D CT scans, over six different datasets achieving state-of-the-art performance, whilst permitting continual learning and avoiding catastrophic forgetting.

Following the training of RACNet, the outputs of the dense layer, consisting of, say L neuron outputs, are extracted for further analysis, employing clustering techniques. These latent variables encapsulate high-level semantic information, integral for producing the ultimate classification at the output layer. Opting to bypass the output layer, unsupervised analysis of these variables is undertaken, aiming to generate a representation set of, say M anchors, which  offers additional insights into the attained decision-making capabilities.

This latent variable extraction and anchor generation by RACNet can be used to assist COVID-19 diagnosis of new 3-D CT scan inputs in an efficient way. The new input is processed through RACNet and the respective vector of  L latent variables is extracted. Then, M distances are computed between this vector and the M anchors (which are also L-dimensional vectors) and the minimum value of these distances is selected, to define the anchor representation that is closest to the examined new case. Moreover, by computing the respective cluster radii, we can provide a confidence level for RACNet's decision, in addition to the confidence levels provided by the RACNet output layer.

The advantage of generation and use of the anchor set is the insight it introduces into the diagnosis process. In each new test case, the generated decision is accompanied by the information about the anchor to which this case was assigned through the above nearest neighbour classification procedure. As a result, the patient, or the doctor, can see which part of RACNet data-driven knowledge was used to make the specific diagnosis. This is crucial for the diagnosis, because, apart from the infection probability, the developed system provides the slices of the anchor CT scan which are more similar to the specific patient's CT-scan that is examined. As a result, this approach enhances the trustworthiness of our system.

Integrating this approach into RACNet  brings an additional advantage.  RACNet  requires retraining with new datasets whenever they become available in different hospitals or medical centres. Privacy concerns often hinder the direct sharing of datasets among medical organizations for retraining purposes. However, it becomes feasible for distinct organizations to share, or discover, high-performing networks trained on external data, along with the corresponding anchor-derived information, via platforms like GitHub. By continually aggregating  anchor sets and associated trained RACNet networks, a collaborative effort can yield shared, enriched data-driven representations that benefit all participants and make COVID-19 diagnosis scalable.

\section{MLOps Orchestration}

\begin{figure} [h!]
\centering
\includegraphics[width=1.\linewidth]{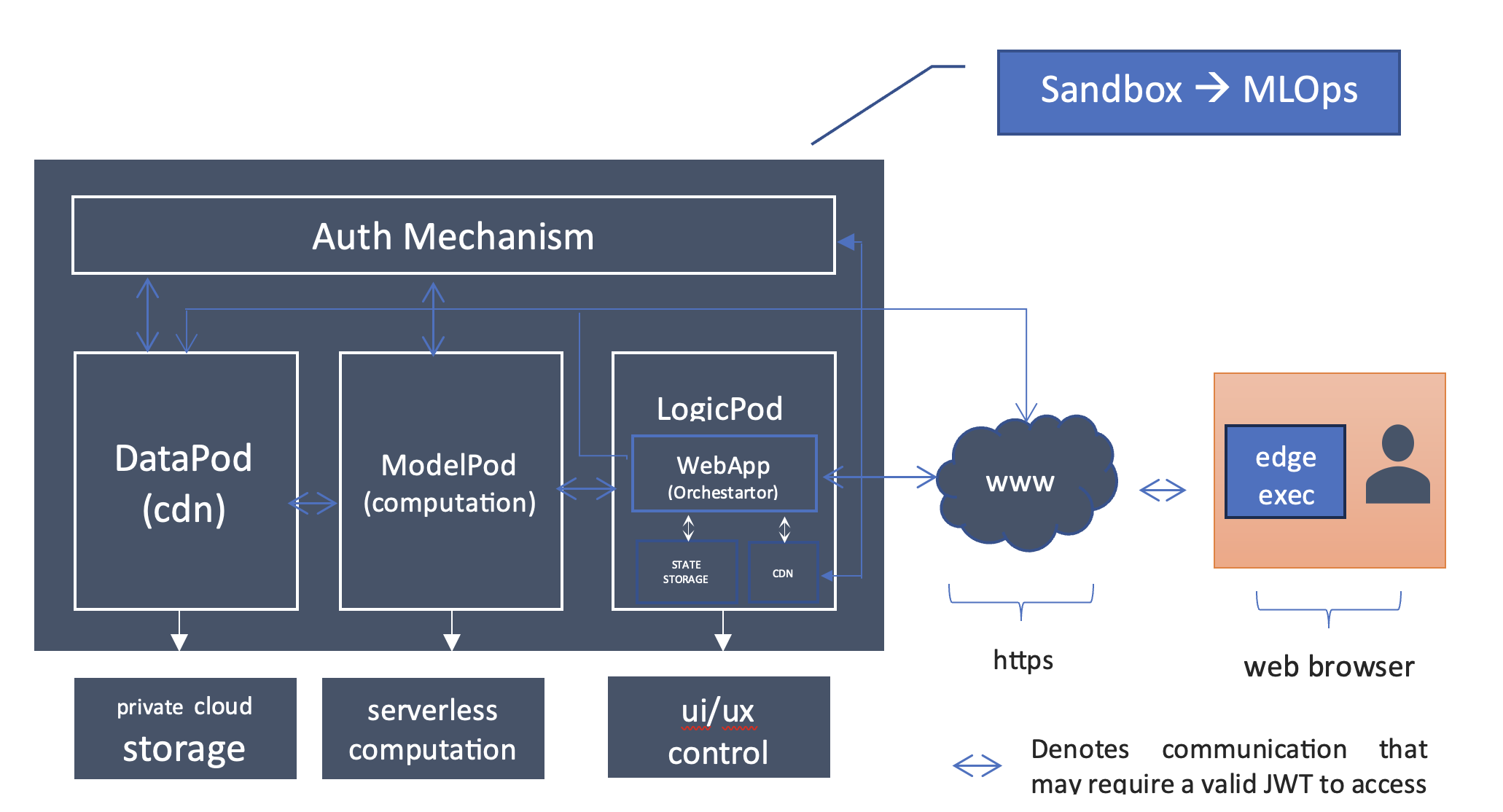}
\caption{A complete diagram that explains the MLPod™ architecture and how the information flows between the various modules. Each module represents an individual service, while the rows visualize the flow of information. Please note that all communications are protected via encryption. The whole deployment of the MLPod™ platform can be considered as an MLOps sandbox.}
\label{fig:mlpod-architecture}
\end{figure}

In the following we present the MLOps orchestration of the RACNet model, which was facilitated by a proprietary framework we developed – MLPod™ \footnote{MLPod™ is a proprietary platform powered by AIandMe SMPC.}. MLOps is a holistic approach that can be extended and be  generally used  for  ML apps deployment. The platform is built on a microservices basis, offering four basic services: i) authentication mechanism, ii) data hosting and sharing, iii) model hosting and execution and, iv) UI/UX. The overall architecture and information flow is depicted in Figure \ref{fig:mlpod-architecture}.

Each subsystem – referred as Pods - deployed to support the corresponding functionality is built as a separate docker deployment, while the service discovery mechanism is implemented as part of the logic microservice (UI/UX), where the corresponding installation for the actual  RACNet model is pre-populated with the required namespace. In brief, the functionality and operation of each individual subsystem of the MLPod™ architecture is described below.

\textbf{Auth Mechanism:} This module handles all the authorization operations. It is capable of issuing and verifying access tokens, following the OAuth 2.0 standard \cite{oauth2}. The tokens supported by the MLPod™ architecture control the access to hosted data, RACNet model execution and the overall access to the deployed Webapp. Each access token may contain information regarding the access permissions (e.g., VPN access restriction, resource allocation restrictions etc.).

\textbf{DataPod:} This module acts as the data storage facility of the whole ML app deployment. Using access tokens, it is responsible for sharing the requested data. In our case, this module holds the information regarding the Covid19 clusters and the representative images and metadata.

\textbf{ModelPod:} This module handles the execution of the DICOM anonymization and the RACNet based Covid19 detection. Using the appropriate access tokens, it authorizes the access to the model execution resources. Furthermore, the core implementation provides a built-in mechanism for supporting both cloud and edge execution of the model. To that end, if the discovery mechanism detects an edge execution request, the ModelPod module automatically wraps the requested model performing the appropriate encryption and validation process and dispatches the model to the edge environment for local execution. We used this functionality  for the DICOM anonymization process so that, any sensitive information (i.e., the related DICOM tags containing the patient’s name, etc.) is removed before it leaves the local environment (i.e, doctor’s working station).

\textbf{LogicPod:} This module is responsible for performing the required orchestration in the application logic. It provides a mechanism to produce the appropriate UI/UX instantiated as a Webapp, via which the end user is accessing the underlying models. LogicPod relies on a proprietary markup language – the \textbf{M}achine \textbf{L}earning \textbf{M}arkup \textbf{L}anguage or ML2 – an XML document that defines the required data inputs, models, and ML pipelines to execute. The LogicPod module can render a ML2 script and convert it to a functional Webapp. Furthermore, it may also contain information for the appropriate service (model) discovery and the execution environment (cloud and/or edge). One may claim that the LogicPod module is a Multimodal AI application deployment and acts as the gateway orchestrator between end users and the underlying models and tasks, i.e. anonymization, diagnosis. The LogicPod module controls the flow of information and model prediction/inference and returns the response to the end user in a human meaningful manner. In particular, the deployed ML2 script instructed the deployed LogicPod module to print the  Covid19 diagnosis provided by RACNet at inference, as a graphical report with texts and images representing the RACNet model based diagnosis and the corresponding explanation of the model’s decision. It should be added that, in case of retraining the model, the new parameters are updated in the LogicPod deployment, without affecting the rest of the application's functionality. The updated model becomes instantly available to the end user  without any need of updating.

\begin{figure} [h!]
\centering
\includegraphics[width=1.\linewidth]{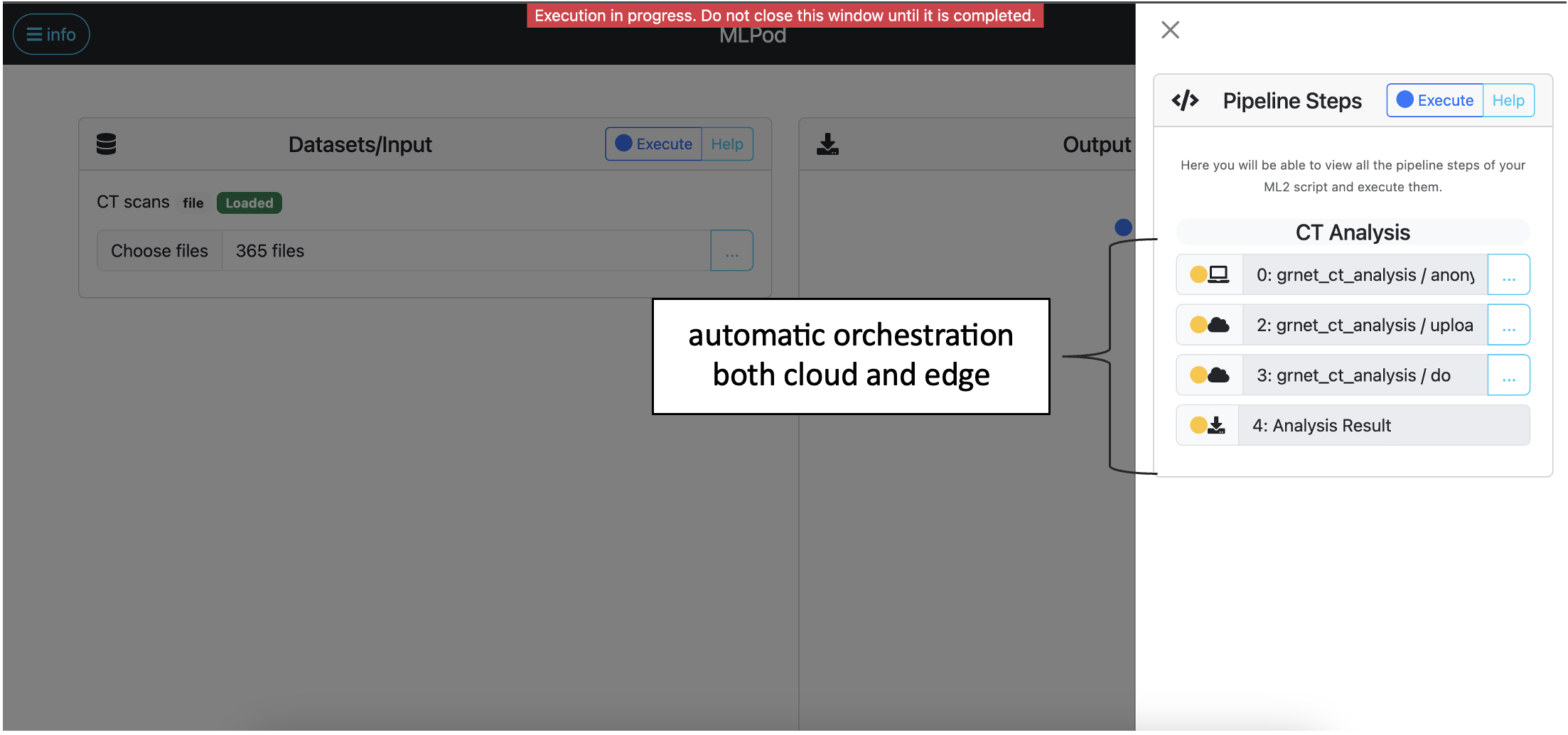}
\caption{A screenshot from an execution of the RACNet based Covid19 detection app. Note the visualization of the underlying pipeline steps. The LogicPod orchestrator automatically handles the communication of the required data in each step. }
\label{fig:app2}
\end{figure}

\begin{figure} [h!]
\centering
\includegraphics[width=1.\linewidth]{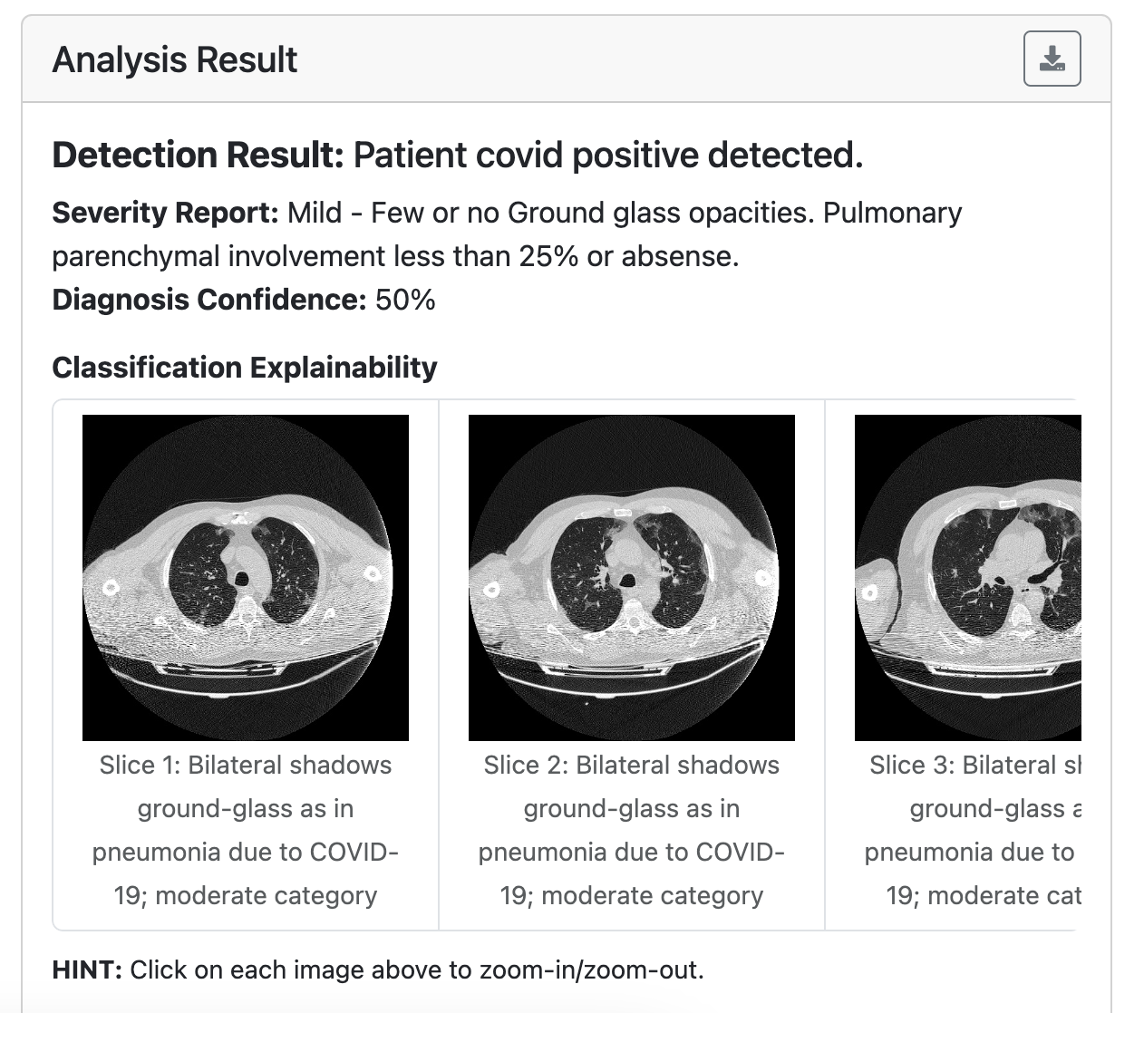}
\caption{An instance of the final response of the RACNet model based diagnosis app. Once the underlying models’ execution is completed, LogicPod module automatically renders the models’ inference and provides a visual report about the patient’s condition, i.e., if the patient is Covid19 positive/negative along with a short text explanation and representative images associated with his/her medical condition.}
\label{fig:app4}
\end{figure}

An example of the deployed RACNet based Covid19 detection MLPod™ is depicted in Figures \ref{fig:app2},\ref{fig:app4}, with representative screenshots to explain the end user’s (doctor) journey.

\section{Compliance with ethical standards}
\label{sec:ethics}

Regarding compliance with ethical standards, 
the paper presents a numerical simulation study and DL-based diagnosis system implementation, for which no ethical
    approval was required. 
Reference is made to the COVID-19 dataset which was developed within former collaboartions, acknowledged in Section 6. 


\vfill
\pagebreak

\section{Conclusion and Future Work}

In this paper we have demonstrated a system architecture for fast, secure and scalable deployment of AI applications on heterogeneous computational environments relying on both cloud and edge execution. Based on that architectural design, the RACNet Covid-19 detection model was built, providing end users (doctors) with an end-to-end user-friendly interface for uploading DICOM images and receiving the diagnosis accompanied by explanation that justifies RACNet's decision. 

To complete the MLOps path, the user's feedback regarding the performance of the model has to be collected and used later for retraining the model and improving its performance. This final step of the MLOps journey is currently being developed and is part of our future work plan.

\section{Acknowledgments}
\label{sec:acknowledgments}

We would like to thank GRNET, National Infrastructures for
Research \& Technology, for supporting us through Project
‘‘Machaon – Advanced networking and computational services to
hospital units in public cloud environment”.The research has been also assisted by the  collaboration between AHEPA Hospital (Radiology Clinic) and
GRNET. The research work of Mr. Arsenos was supported by the Hellenic Foundation for Research
and Innovation (HFRI) under the 3rd Call for HFRI PhD Fellowships    
(Number: 4941/23).

\bibliographystyle{IEEEbib}
\bibliography{strings}

\end{document}